\documentclass[prd,12pt,nofootinbib]{revtex4-2}
\usepackage{tikz-cd}
\usepackage{amssymb,amsmath,braket,comment}

\DeclareSymbolFont{usualmathcal}{OMS}{cmsy}{m}{n}
\DeclareSymbolFontAlphabet{\mathcal}{usualmathcal}
\def\KD{K\"{a}hler-Dirac }

\def\bx{{\bf x}}

\def\ha{{\bf\hat{a}}}

\newcommand{\BZ}{{\mathbb Z}}

\def\cT{{\cal T}}

\begin{document}

\title{Supersymmetric lattice theories on curved space}
\author{
David Berenstein \textsuperscript{1} }
\author{
Simon Catterall \textsuperscript{2}} 

\affiliation{\textsuperscript{1}Department of Physics, University of California, Santa Barbara, CA 93106, USA}
\affiliation{\textsuperscript{2}Department of Physics, Syracuse University, Syracuse, 13244, New York, USA}

\begin{abstract} We show how to construct Hamiltonian lattice theories
with one exact supersymmetry on arbitrary triangulations of curved space in any
number of dimensions. Both bosons and fermions satisfy discrete
K\"{a}hler-Dirac equations.  The quantization of the fermions proceeds by
imposing conventional anti-commutation relations while the bosons
require a modification of the usual canonical commutator. On regular lattices we construct parity, time reversal and translation-by-one (shift) symmetries. We argue that the latter are generically non-invertible symmetries. We also show how to couple these degrees of freedom to background gauge fields which leads to a theory with enhanced supersymmetry.
\end{abstract}

\maketitle
\section{Introduction}

Staggered fermions are familiar from efforts in lattice gauge theory to
handle the fermion doubling problem \cite{Kogut:1974ag}. Their generalizations to curved space are called \KD fermions. In recent work it has been shown using path
integral methods that these theories suffer from
new gravitational anomalies that survive discretization \cite{Catterall:2018lkj,Catterall:2018dns,Butt:2021brl,Catterall:2022jky,Catterall:2023nww}. They also have Hamiltonian formulations which have been studied
in flat space and exhibit mixed lattice 't Hooft anomalies involving translation-by-one or shift symmetries and parity or time reversal symmetry \cite{Seiberg:2023cdc,Li:2024dpq,Gioia:2025bhl,Chatterjee:2024gje,Pace:2025rfu,Onogi:2025xir}.

In this paper we focus on a different aspect of
these theories - namely that they can be supersymmetized on any random triangulation. In this way they can be coupled to gravity while preserving a global
supersymmetry. These constructions
go beyond the supersymmetric lattice theories that have been studied before which were restricted to flat space and
regular lattices \cite{Catterall:2009it,Catterall:2023tmr}. 

We will see that the boson fields in this construction
also satisfy the \KD equation. Usually this would be incompatible
with the spin-statistics theorem since this would seem to imply
a bosonic action that is first order in derivatives. In our construction, however, the bosonic Hamiltonian contains no derivatives and instead the unusual equation of
motion arises from a modification of the canonical commutation relations. 
In flat space the bosonic sector corresponds to a theory of {\it staggered bosons} 
which have also been studied recently in connection to non-invertible symmetries \cite{Berenstein:2023tru,Berenstein:2023ric}. It is expected that theories with these types of symmetries have interesting anomalies - see \cite{Shao:2023gho} for a review.

In section II we develop the simplest version of the staggered boson construction
in one spatial dimension, examining the role of parity and time reversal
symmetries and showing how to construct a scalar supersymmetry
that generates a one dimensional staggered fermion. In section III we
generalize this construction to any number of flat space dimensions. In section IV we give the corresponding formulation for arbitrary random triangulations where the staggered
fields are replaced by \KD fields. Finally we summarise our conclusions in
section V and outline further work.

\section{One flat spatial dimension}
\subsection{Staggered bosons}

Our bosons denoted $q_\alpha$ live on the sites $\alpha$
of a one dimensional lattice with $n$ sites. In the absence of
independent canonical momenta we cannot quantize these boson
operators with canonical
commutation relations. Instead we will impose 
the commutation relation 
\begin{equation}
    [q_\alpha,q_\beta]=is(\alpha,\beta)\label{eq:comm1}
\end{equation}
where 
\begin{equation}
    s(\alpha,\beta)= \eta(\alpha,\beta)\delta_{\alpha+1 ,\beta}-\eta(\alpha ,\beta)\delta_{\alpha-1, \beta} 
\end{equation}
where $\eta(\alpha,\beta)$ is a $\BZ_2$ valued link variable. Thus the commutator
is only non-zero for fields on nearest neighbor sites.
It is convenient for us to introduce a graphic representation of equation \eqref{eq:comm1} as follows.
Whenever $s(\alpha,\beta)=1$ we draw an arrow from site $\alpha$ to site $\beta$. When $s(\alpha,\beta)=-1$ we draw an arrow in the opposite direction. The commutator eqn.~\ref{eq:comm1} implies that $s(\alpha,\beta)$ must be
antisymmetric. This restricts the possible assignments of the $\eta$ or the arrows.
\begin{figure}[ht]
\begin{center}
    \begin{tikzcd}
    \bullet \arrow[r] &\bullet \ar[r]  & \bullet& \dots \ar[r]& \bullet\\
    0 & 1 &2 & \dots &n\equiv 0
    \end{tikzcd}\caption{Periodic staggered boson in 1D}\label{fig:periodic1}
\end{center}
\end{figure}
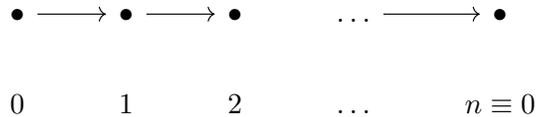
For example, a periodic staggered boson lattice system in one dimension will be represented as in figure \ref{fig:periodic1} and has all $\eta=1$.

We will also consider the simplest possible Hamiltonian:
\begin{equation}
    H= \sum_{\beta=0}^{n-1} \frac 12 q_\beta^2\label{simpleH}
\end{equation}
The Heisenberg equations of motion yield
\begin{equation}
    \dot q_{\alpha} = i [H, q_{\alpha}]= ( q_{\alpha+1}- q_{\alpha-1})
\end{equation}
which looks like a discretization of a first order wave equation $\partial_t \phi= 2 \partial_x \phi$ for a left moving field in $1+1$ dimensions. We can immediately solve the equations of motion using a plane wave decomposition, $q_r= a_k(t) \exp(i 2\pi k r/n) $. We find
\begin{equation}
   \dot a_k(t)=  (\exp(i 2\pi k/n)-\exp(-i 2\pi k/n))a_k(t)=2i \sin(2\pi k/n) a_k(t) 
\end{equation}
The solution $a_k(t)= a_k(0)\exp(-i \omega(k) t)$ has a frequency
$\omega(k)= -2 \sin(2\pi k/ n)$.
The modes with $\omega\neq 0$ are regular harmonic oscillators, where the pair $a_{-k}, a_{k}$ are a pair of raising/lowering operators.~\footnote{From
$a_k\propto  \sum_r \exp(-2\pi i k r/n) q_r$ it is easy to see that $a_k^\dagger= a_{-k}$.}
We will call $p=2\pi k/n $ the quasi-momentum of the mode. 

It is important
to notice that the theory is invariant under a $\BZ_2$ gauge
transformation 
\begin{equation}
    q_\alpha\to \phi_\alpha q_\alpha\quad{\rm and}\quad \eta_{\alpha\beta}\to\phi_\alpha\eta_{\alpha\beta}\phi_\beta\quad \phi_{\alpha,\beta}\in \BZ_2
\end{equation}
In the graphical notation, sending $q_\alpha\to -q_\alpha$ reverses the arrows at lattice site $\alpha$. A global transformation can be composed of $n$ such elementary moves. 

The system possesses additional global symmetries. For example we can
define a parity operation that exchanges the sites $\alpha\to n-\alpha$.
This flips $s(\alpha,\beta)\to -s(\alpha,\beta)$. It can be
rendered a symmetry {\it when $n$ is even} by a field transformation
\begin{equation}
P: q_k \to (-1)^k q_{n-k}
\end{equation}
The graphical proof is as follows. The parity operation that just exchanges $q_\alpha\to q_{n-\alpha}$ redraws the drawing in figure \ref{fig:periodic1} with  opposite labels. When restoring the correct order  it is clear all arrows are reversed. 
We can undo that by a gauge transformation if $n$ is even, as we can flip the sign of all the odd sites and this will change the direction of all the arrows.
It is only possible to divide the sites into even and odd sites when $n$ is even, if we are also imposing periodic boundary conditions.

Notice that this sends the mode of momentum $a^\dagger_p \to a^\dagger_{\pi-p}$ where we have employed $(-1)^k\equiv \exp(i \pi k)$. The modes near zero are left movers, the modes near $\pi$ are right movers. The parity operation sends left movers into right-movers and vice versa. This parity symmetry is broken if $n$ is odd. 
This can be seen as follows.
If we do the same graphic parity operation as before, we can change the sign of most arrows except the last one.
This ends up producing a different topological sector.

Indeed, if $n$ is even and we have one arrow in the wrong direction we have effectively imposed antiperiodic boundary conditions. 
In fact this is equivalent
to the presence of a non-trivial $\BZ_2$ background field - one link field
$\eta=-1$ while
all others are taken to be unity. The holonomy of the $\BZ_2$ gauge field on the circle is then $-1$ , which is clearly invariant under gauge transformations. We will call this a $\BZ_2$ twisted sector. It is 
also invariant under parity.
At the level of quasi-momentum $p$,
$p$ is quantized in units of $2\pi/n$. In the twisted sector at $n$ even, there is an extra shift of $\pi/n$ in the momentum labels. 

It is easy to see that when $n$ is odd there is a mode $a_0$ at zero frequency, but there is no $a_\pi$ at zero frequency. Under the parity operation, we go from an untwisted sector to a $\BZ_2$ twisted sector. We now have a mode $a_\pi$ at zero frequency, but no mode $a_0$.

Another important symmetry when $n$ is even is time reversal ${\cal T}$ which reverses the time coordinate and changes the $i\to -i$. If 
\begin{equation}
    q_\alpha\stackrel{{\cal T}}{\to}-\left(-1\right)^\alpha q_\alpha
\end{equation}
the equation of motion and 
Hamiltonian are left unchanged under this operation.

Finally both the Hamiltonian and the
equations of motion
are invariant under a translation-by-one or {\it shift
symmetry} $S$:
\begin{equation}
    q_\alpha\stackrel{S}{\to}\xi(\alpha,\alpha+1)q_{\alpha+1}
\end{equation}
where $\xi\in \BZ_2$.
Demanding this be an invariance requires that
\begin{equation}
    \xi(\alpha-1,\alpha)\eta(\alpha-1,\alpha)\xi(\alpha,\alpha+1)=\eta(\alpha,\alpha+1)
\end{equation}
In the gauge where $\eta=1$ this can be satisfied by $\xi=1$.
Notice that this symmetry does not commute with either time reversal or parity
so that that it should not be interpreted as lattice translation. Lattice translations
instead correspond to translation by two sites. 
The fact that shift symmetries do not commute with other discrete spacetime symmetries suggests that these symmetries have non-trivial mixed anomalies.
Finally it should be noted that with periodic boundary conditions
the system possesses two conserved charges (which are also modes of zero frequency)
\begin{align}
    C_{e}&=\sum_{i\;{\rm even}} q_i\nonumber\\
    C_{o}&=\sum_{i\;{\rm odd}} q_i
\end{align}
By our conventions, ${\cal T}: C_{e}\to -C_{e}, C_{o}\to C_{o} $, whereas parity sends $C_{e}\to C_{e}$, $C_{o}\to -C_{o}$. Alternatively one can define 
\begin{align}
    C_\pm=C_e\pm C_o
\end{align}
which are eigenstates of parity and time reversal.
The full Hilbert space of the theory is given by
a sum over superselection sectors each of which corresponds to fixed values for
$C_{e}$ and $C_o$. However, notice that the shift symmetry mixes between superselection sectors and
hence will only remain a good symmetry on a subset of Hilbert
space. This restriction means that the
operator representing the shift needs to be augmented by appropriate
projectors if they are to remain true symmetries of the system. The bosonic
shifts should hence be thought of as non-invertible symmetries - see \cite{Berenstein:2023ric} for
an extended discussion.

\subsection{Supersymmetry and staggered fermions}

To our system of bosons we
will now add one (real) fermion at each site $\theta_\alpha$.
This follows ideas from our previous work \cite{Berenstein:2024tdc}. We will posit
that these satisfy the usual anti-commutation relation:
\begin{equation}\{\theta _\alpha, \theta_\beta\}= 2 \delta_{\alpha,\beta}
\end{equation}
We will assume that $[\theta_\alpha,q_\beta]=0$.
We can then
declare that the following fermionic, self-adjoint operator $Q$ is a supersymmetry
\begin{equation}
    Q= \sum _\alpha \frac 1{\sqrt{2}} q_\alpha \theta_\alpha
\end{equation}
The corresponding Hamiltonian is $H=Q^2$. A straightforward computation shows that 
\begin{eqnarray}
    Q^2&=& \sum_\alpha \frac 12 q_\alpha^2+ \frac{1}{2} \theta_\alpha \theta_{\alpha+1}[q_\alpha,q_{\alpha+1}]\\
    &=& \sum_\alpha \frac 12 q_\alpha^2+ \frac{i}{2} \theta_\alpha \theta_{\alpha+1}
\end{eqnarray}
so $Q^2$ gives us back the Hamiltonian for the staggered bosons $q$, while for the fermions it produces a Majorana chain. The latter is in fact nothing more than
a (massless) reduced staggered fermion in one dimension. This is most easily seen by using the algebraic identity 
\begin{equation}
 \theta_\alpha\theta_{\alpha+1} =\frac{1}{2}\theta_\alpha\left(\theta_{\alpha+1}-\theta_{\alpha-1}\right)
\end{equation}
Using the supersymmetry, we also have the relation 
\begin{equation}
q_\alpha= (\sqrt{2})^{-1} \{ Q, \theta_\alpha\}  
\end{equation} 
and it is easy to see that the equations of motion of the $\theta_\alpha$ are the same as those of the $q_\alpha$. So the spectrum of the (one particle) free fermion and the spectrum of the (one particle) boson is the same.

We can also check that the two bosonic zero modes can be obtained as follows
\begin{equation}
    C_\pm\propto \{Q, \eta_\pm\}
\end{equation}
so we also have two fermion zero modes $\eta_\pm$ with well defined
parity which will
mix under the shift symmetry. This suggests that a shift should be interpreted as 
a discrete axial transformation in the continuum limit.
Because the zero modes anticommute and square to a c-number, we can represent them by Pauli matrices $\eta_+\propto \sigma_x, \eta_-\propto \sigma_y$. There are then two possible ground states distinguished by the eigenvalue under $\gamma_5=\sigma_z=\eta_+\eta_-$. Multiplying any ground state by $\eta_-$ or $\eta_+$ changes the eigenvalue of $\gamma_5$ while a shift swaps the two ground states, taking $\gamma_5\to -\gamma_5$. 

These ideas can be lifted to any dimension as we show in the next section.
An in depth discussion of the fermion symmetries associated to this one dimensional chain and its relation to non-invertible symmetries can be found in \cite{Seiberg:2023cdc}.

\section{More than one flat dimension}
\subsection{Staggered bosons}

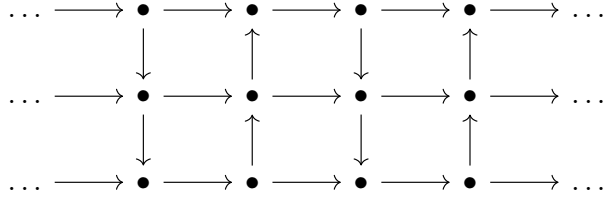
\begin{figure}[ht]
\begin{center}
    \begin{tikzcd}
   \dots \ar[r]&\bullet \arrow[r] \ar[d] &\bullet \ar[r ]  & \bullet \ar[r] \ar[d] &\bullet \ar[r ] &\dots\\
   \dots \ar[r]  & \bullet \arrow[r] \ar[d] &\bullet \ar[r ] \ar[u] & \bullet \ar[r] \ar[d] &\bullet \ar[r ] \ar[u]&\dots\\
 \dots \ar[r] &  \bullet \ar[r] &\bullet \ar[r ] \ar[u]&\bullet \ar[r ] &\bullet \ar[r ]\ar[u] &\dots
    \end{tikzcd}\caption{Relativistic staggered boson algebra in 2D}\label{fig:z2plaquette}
\end{center}
\end{figure}

Let's see how the one dimensional construction can be generalized to lattices in higher dimensions.  We again assume a single real boson $q$ on each site of a regular
hypercubic lattice equipped with periodic boundary conditions.
The one dimensional commutator   becomes
\begin{equation}
[q_{\vec \alpha},q_{\vec\beta}]=is(\vec \alpha,\vec \beta)
    \label{comm}
\end{equation}
where $s$ is given by 
\begin{equation}
    s(\vec \alpha,\vec \beta)= \sum_a  \left[\eta_a(\vec\alpha,\vec\beta)\delta_{\vec \alpha+e_a ,\vec \beta}-\eta_a(\vec\alpha ,\vec\beta)\delta_{\vec \alpha-e_a,\vec \beta}\right]  \label{eq:comm_full}
\end{equation}
where $\eta_a(\vec{\alpha})$ is again a $\BZ_2$ valued gauge field associated with
a link $\alpha\to \alpha+e_a$ where $e_a$ is a unit vector in the lattice. There
is one crucial new feature that appears for dimensions greater than one namely that
if we want to obtain a relativistic continuum theory we need to
impose the requirement that
any plaquette has ${\mathbb Z}_2$ holonomy minus one. That is
\begin{equation}
\eta_a(\vec\alpha)\eta_b(\vec\alpha+e_a)\eta_a(\vec\alpha+e_b)\eta_b(\vec\alpha)=-1
\end{equation}
One can think of this constraint as arising from the limit of the $\BZ_2$ gauge
theory when
the lattice gauge coupling $\beta\to -\infty$. 
If we impose
the additional (gauge fixing) condition that
we want $\eta_a(\vec\alpha)$ to be constant along the direction $e_a$ i.e
the Lorentz gauge $\Delta_a\eta_a(\alpha)=\eta_a(\alpha)-\eta_a(x-e_a)=0$ we can write down a solution to these constraints 
\begin{equation}
    \eta_a(\vec \alpha)
= (-1)^{\sum^{a-1}_{i=1}\alpha_i}\label{eq:parity function}
\end{equation}
As an example, in our graphical notation, for a $2+1$ dimensional field theory we would find
the algebra is represented by figure \ref{fig:z2plaquette} (up to field redefinitions).
The phase in \eqref{eq:parity function} is precisely the
phase  that arises in staggered lattice fermions \cite{Susskind:1976jm,Golterman:1984cy}.

The Hamiltonian is again taken to be
\begin{equation}
    H=\frac{1}{2}\sum_{\vec \alpha} q^2_{\vec \alpha}
\end{equation}
This leads to the equation of motion
\begin{align}
    \frac{\partial q_{\vec\alpha}}{\partial t}&=i[H,q_{\vec\alpha}]\nonumber\\
    &=\frac{i}{2}\sum_{\vec \beta} \left(q^2_{\vec \beta}q_{\vec \alpha}-q_{\vec \alpha}q^2_{\vec \beta}\right)\nonumber\\
    &=\sum_{\beta} s(\vec \alpha,\vec \beta) q_{\vec \beta}
    \label{eom}
\end{align}
and we see by expanding $s(\vec \alpha,\vec\beta)$ that we sum over neighbors along all $e_a$ with alternating signs and phases $\eta_a(\vec \alpha)$.
Using the result that every plaquette should be minus one and that the $\eta_a$ phase is also translation invariant along $e_a$, it easily follows that
\begin{equation}
    \eta_a(\vec \alpha )\eta_b(\vec \alpha+e_a)+\eta_b(\vec \alpha)\eta_a(\vec \alpha+e_b)=2\delta_{ab}.
\end{equation}
One can now easily show that the field $q$ satisfies a discrete wave equation in arbitrary dimensions
\begin{equation}
    \frac{\partial^2 q_{\alpha}}{\partial t^2}=\sum_a\left(q_{\vec \alpha+2 e_a}  +q_{\vec \alpha-2e_a}-2q_{\vec\alpha}\right)
\end{equation}
Notice the appearance of a discrete Laplacian operator on a block lattice with twice the lattice spacing. This is important in that it shows that the system produces the same relativistic dispersion relation for every element of a multiplet of size $2^d$ corresponding
to the points in the unit cell of the block lattice. These can be labeled
by the parity of each coordinate $\alpha_i$. 
We will call the collections of vertices of a fundamental hypercube of the lattice a block. The set of translations by two units in each direction on a fundamental block covers 
the lattice.

Let us now consider the effects of translations of the fields by one lattice spacing rather than two - the shift operation we discussed earlier.  For example, an elementary shift $S_a$ of the field in direction $e_a$ acts as 
\begin{equation}
    q_{\alpha}\stackrel{S_a}{\to} \xi_a(\vec \alpha)q_{\vec\alpha+e_a}
\end{equation}
We want this translation to be an automorphism of the algebra defined by \eqref{eq:comm_full}, so $\xi$ will be a gauge transformation (a sign) that we need to adjust to get translation invariance. For the explicit choice of
phases given in eqn.~\ref{eq:parity function} it is not hard to see that we get
a symmetry if we adopt the definition
\begin{equation}
\xi_a(\vec\alpha)=\left(-1\right)^{\sum_{i=a+1}^d \alpha_i}
\end{equation}
This is precisely the usual
phase associated with shift symmetries in staggered fermion theories see eg. \cite{Susskind:1976jm, Golterman:1984cy}. It is conjugate to $\eta_a(\vec\alpha)$ in the sense that
\begin{equation}
    \xi_a(e_b)\eta_b(e_a)=1\label{eq:gauge_cond}
\end{equation}
Notice that $\xi_a(\vec\alpha\pm e_a)=\xi_a(\vec\alpha)$.

Furthermore, it should be clear
that these simple shifts can be applied consecutively generating more
symmetries. For example
the double shift $S_{ab}$
\begin{equation}
    q_{\vec \alpha}\stackrel{S_{ab}}{\to} S_a S_b\, q_
    {\vec\alpha}=\xi_b(\vec \alpha)\xi_a(\vec\alpha+e_b)q_{\vec\alpha+e_a+e_b}\quad{\rm where}\quad a\ne b
\end{equation}
Using the property $\xi_a(\vec \alpha +e_b)=-\xi_b(\vec \alpha+e_a)$ for $e_a\ne e_b$ it is trivial to see that $S_a S_b=-S_b S_a$.
That is, translations by one anticommute.

Notice that a double shift along the same direction yields a simple translation $T$ on the block lattice.
\begin{equation}
    q_{\vec\alpha}\stackrel{S_a^2}{\to}\xi_a(\vec\alpha)\xi_a(\vec\alpha+e_a)q(\vec \alpha+2e_a)=q_{\vec \alpha+2 e_a}=T_{2 e_a}\left[q_{\vec \alpha}\right]
\end{equation}
essentially because $\xi_a$ is translation invariant along $e_a$.

Performing a $S_{ab}$ shift followed by a $S_{b}$ shift reveals a non-trivial algebra involving both shifts and translations where the combination of two shifts generates another shift up to a block lattice translation $T$. In fact, it is not hard to see
that in dimension $d$ there are $2^d$ possible shifts which are in
one to one correspondence with the elements of the Clifford algebra $\Gamma_d=\left(I,\gamma_a,\gamma_a\gamma_b,\ldots\right)$.
\begin{equation}
    q(\vec{\alpha})\to \Xi_A(\vec \alpha)q_{\vec \alpha +e_{A}}
\end{equation}
where the vector $e_A$ runs over all shifts formed out 
of successive elementary shifts and the phase $\Xi_A(x)$ is the corresponding product of elementary shift phases:
\begin{align}
     e_A&=\sum e_a+e_b+\cdots\\
     \Xi_A&=\prod \xi_a(\vec \alpha)\xi_b(\vec \alpha+e_a)\cdots
\end{align}
If the $\Xi_A$ are thought of as Clifford algebra elements, the natural interpretation is that they are objects that act on bosonic k-form valued fields.

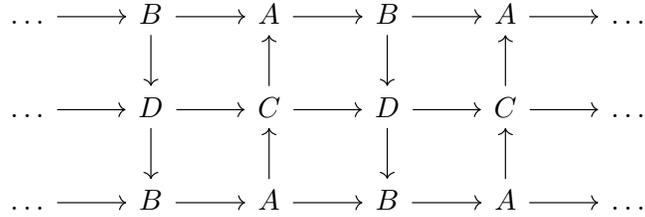
\begin{figure}[ht]
\begin{center}
    \begin{tikzcd}
   \dots \ar[r]&B \arrow[r] \ar[d] &A \ar[r ]  & B \ar[r] \ar[d] &A  \ar[r ] &\dots\\
   \dots \ar[r]  & D \arrow[r] \ar[d] &C \ar[r ] \ar[u] & D \ar[r] \ar[d] &C \ar[r ] \ar[u]&\dots\\
 \dots \ar[r] &  B \ar[r] &A \ar[r ] \ar[u]&B \ar[r ] &A \ar[r ]\ar[u] &\dots
    \end{tikzcd}
\end{center}\caption{Four coloring of sites for center}\label{fig:fourcoloringABCD}
\end{figure}

To understand this better, we need to look at how the fields transform under rotations of the lattice. Let us take for example the $2+1$ field theory and the coloring of vertices according to their possible vertex labels modulo $2$ as in the figure \ref{fig:fourcoloringABCD}.
Take a vertex with label $A$ and consider a $90^\circ$ rotation of the lattice theory about that vertex $A$. If we declare that the field at $A$ is a scalar $q_A\to q_A$, we see that other vertices with the $A$ label get rotated into vertices with the same label. However, $B$ gets rotated into $C$ and $C$ into $B$ and $D$ gets rotated into $D$. When we do the rotations the arrows change directions, so that $q_B\to q_C, q_C\to -q_B, q_D\to -q_D$. We can check that these same signs appear in the rotations of the different k-forms $dx, dy, dx\wedge dy$. That makes the identification with k-forms natural. 

The map in general dimensions is the following. First, we need to declare that one of the sites is the origin. There are $2^D$ labels, obtained after translation by one in multiple directions $e_{a_1}+\dots +e_{a_k}$ where the $a_i$ are strictly increasing. The fields with such labels are assigned the form $e_{a_1}\wedge \dots e_{a_k}$. It can be checked that rotations are consistent with this labeling, so that the label and the form type coincide. Much of this will become even more evident when we discuss the \KD equation later which naturally operates
on forms.

As earlier we will be interested in the zero modes of this system, namely fields
$q_{\vec{\beta}}$ that are annihilated by the matrix $s(\vec{\alpha},\vec{\beta})$.
One of these trivially corresponds to the constant field $q_{\vec{\alpha}}=1$.
But there are in fact $2^d-1$ additional zero modes generated by the shifts
\begin{equation}
    S_A q_{\vec{\alpha}}=\Xi_A(\vec \alpha)q_{\vec \alpha+e_A}=\Xi_A(\vec \alpha)
\end{equation}
Associated with these zero modes there are a set of charges $C_A$ that commute with $H$ and are hence conserved $\frac{dC_A}{dt}=0$
\begin{equation}
    C_A= \sum_{\vec{\alpha}} \Xi_A(\vec\alpha) q_{\vec \alpha}
\end{equation}

A nice way to write $C_A$ is to notice that they can be gotten by adding up
the contributions of $q_{\vec{\alpha}}$ from sites whose coordinates carry the same parities as the shift vector $\vec{A}$ i.e that satisfy
the condition $\alpha_i\;{\rm mod}\;2=A_i$ where $A_i$ are
the coordinates of the shift vector $\vec{A}$. Such lattice sites span
a sublattice $L_A$
\begin{equation}
    C_A = \sum_{\vec \alpha \in L_A} q_{\vec \alpha}
\end{equation}
It is also easy to see that $[C_A,C_{A^\prime}]=0$.

However, notice that although
the Hamiltonian and equations of motion are invariant under the shifts, this operation clearly again mixes the $C_A$. 
For example $S_a C_A= C_{A\pm e_a}$.
Since the $C_A$ are central, they label irreducible representations of the staggered boson algebra and these can be diagonalized simultaneously with the Hamiltonian. 
What this means is that the shifts are outer automorphisms of the staggered 
boson algebra. Just like in $1+1$ dimensions only a subset of the full Hilbert space
is invariant under these shifts and hence the latter
must, in general, be modified to include suitable
projectors and hence should be treated as non-invertible symmetries. The precise description of the non-invertible symmetries depends on the quantization conditions on the modes $C$. This is beyond the scope of the present article.

Again one can show that the system is invariant under both parity
and time reversal transformation. The latter acts as
\begin{equation}
    q_{\vec \alpha}\stackrel{\cT}{\to}-\epsilon(\vec \alpha)q_{\vec \alpha}\quad t\to -t\quad   i\to -i. \label{eq:timerev}
\end{equation}
where the site parity $\epsilon(\vec \alpha)= (-1)^{\sum_i \alpha_i}$.
It is straightforward to verify that both
the Hamiltonian and equations of motion 
are invariant under $\cT$.

The set of conserved charges $C_A$ thus
breaks into two sets under $\cT$: those arising from odd shift zero modes which are even under $\cT$ 
\begin{equation}
    \cT C_{\rm odd} \cT^{-1}=C_{\rm odd}\end{equation}
and charges derived from even shift zero modes which are odd
\begin{equation}
    \cT C_{\rm even}\cT^{-1}=-C_{\rm even}
\end{equation}
The choice of sign in \eqref{eq:timerev} is a convention that we
choose to fit with our earlier discussion of
the one dimensional case.

However, it is also 
easy to see that, as in one dimension,
odd shifts anticommute with ${\cT}$ \cite{Catterall:2025vrx}. 
\begin{align}
 S_a\cT: q(\bx)&=\epsilon(\bx)\xi_a(\bx)q(\bx+\ha)\nonumber\\   
 \cT S_a: q(\bx)&=\epsilon(x+\ha)\xi_a(\bx)q(\bx+\ha)=-\epsilon(\bx)\xi_a(\bx)q(\bx+\ha)
\end{align}
This suggests that there may be a mixed anomaly between
$\cT$ and $S^{\rm odd}$. 

\subsection{Supersymmetry and staggered fermions in  higher dimensions}

If we introduce a (Majorana) fermionic partner $\theta$ on each site it is easy to construct a supercharge that generates the bosonic Hamiltonian we have considered previously.
\begin{equation}
    Q=\frac{1}{\sqrt{2}}\sum_{\vec \alpha} \theta_{\vec \alpha}q_{\vec \alpha}
\end{equation}
where the fermions satisfy the anticommutation relations $
\{\theta_{\vec \alpha}, \theta_{\vec \beta}\}=2\delta_{\vec\alpha,\vec\beta}$.

Defining $H=Q^2$ we find after a little algebra
\begin{equation}
    H=\frac{1}{2}\sum_{\bx} q_{\vec \alpha}^2+\frac{i}{2}\sum_{\vec\alpha,a}\theta_{\vec\alpha}\eta_a(\vec \alpha) \theta_{\vec\alpha+e_a} 
\end{equation}
 This is nothing more than the Hamiltonian for a reduced staggered
 fermion.~\footnote{The designation {\it reduced}just means there is only one degree of freedom
 per site. It is the analog of a Majorana condition.} In the
 continuum limit it gives rise to $2^{\frac{d-1}{2}}$ Dirac fermions. 
 The bosonic shift symmetries of this theory then imply the existence of corresponding fermionic shift symmetries which are related to flavor symmetries
of the continuum Dirac theory \cite{Catterall:2025vrx}.
These models are very similar to those built in \cite{Elitzur:1982vh,Elitzur:1983nj}, where the fermions of the supersymmetric model are also realized as staggered fermions. In those models, the role of the staggerd bosons would be realized as either momenta, or gradients. They were also able to introduce non-trivial interactions in some cases.

If we insist that $Q$ is invariant under time reversal and parity  then 
these symmetries also hold for the staggered fermion theory. Also, the anomalies that exist for bosonic symmetries  extend immediately to the fermions. Notice that $Q$ is treated as a scalar, so if the usual connection between spin and statistics is used, $Q$ has the wrong spin. The usual way to fix this is to think of $Q$ as a twisted supersymmetry, where the twist is by some internal flavor symmetry \cite{Catterall:2009it}. Then both the fermions and bosons can become $k$-forms in the continuum limit.

\subsection{Coupling to a background gauge field}

We now want to study the case where the fermions carry a $U(1)$ charge (or more generally a $U(n)$ charge).
To do this, we need to promote the fermions to be complex, rather than real. So at each site we will have $\eta_\alpha, \bar \eta_\alpha=\eta_\alpha^\dagger $ with $\{\eta_\alpha,\eta_\beta\}=0$ and $\{\eta_\alpha,\bar \eta_\beta\}= 2 \delta_{\alpha\beta}$.
Supersymmetry suggests that if $Q$ is invariant under the $U(1)$ transformations, then the bosons will also need to become complex.
So how do we insert the gauge field?
The basic idea is to twist the commutator relation by gauge connections on the links as follows
\begin{equation}
[\bar q_\alpha, q_\beta]= i s(\alpha,\beta) U_{\alpha\beta}
\end{equation}
with the understanding that $U_{\alpha\beta}=U_{\beta\alpha}^{-1}$ with the same $s$ function we had above. 
For each $q_\alpha$ we need a complex conjugate variable  with the opposite charge because of the fact that we are dealing with quantum mechanics and we are always allowed to take the adjoint of an operator. 

It is easy to check that the supersymmetry can be split into two pieces,
\begin{equation}
    Q= \frac 1{\sqrt{2}} \sum q_\alpha \bar \eta_\alpha ,\quad  \bar Q= \frac 1{\sqrt{2}} \sum \bar q_\alpha \eta_\alpha
\end{equation}
We clearly have that $Q^2=\bar Q^2=0$ and
it is easy to see that our standard Hamiltonian arises from $H=(Q+\bar Q)^2= (-i(Q-\bar Q))^2$. Therefore $H$ commutes with $Q,\bar Q$ separately and we have doubled the amount of supersymmetry that is directly visible in the lattice. $Q^2$ then produces exactly a gauged version of the \KD action for the fermions.
It is straightforward to generalize to $U(n)$ by having fields $q_A^I$ transforming for example in the fundamental of $U(n)$ and making the $U$ also carry $I,J$ indices.
Notice, however, that this construction does not generate a kinetic term for the gauge fields. This procedure is therefore capable of coupling the theory to
background $U(N)$ gauge fields suitable for probing the 't Hooft anomaly
structure of the theory but cannot be used for understanding the non-perturbative
structure of the theory with dynamical gauge fields.

\section{Formulation on random triangulations}
\subsection{\KD fields and curved space supersymmetry}
This staggered boson construction can be generalized to arbitrary triangulations of curved space~\footnote{Actually arbitrary cellular decompositions are also possible.}.
One simply replaces eqn. \eqref{comm} by
\begin{equation}
    [q^n_\alpha,q^m_\beta]=iK^{n,m}_{\alpha\beta}
\end{equation}
where the \KD operator $K$ is given by
\begin{equation}
    K^{nm}_{\alpha\beta}=\left(\delta_{m,n-1}I^{n,m}_{\alpha\beta}-\delta_{m,n+1}\bar{I}^{n,m}_{\alpha\beta}\right)
    \label{random}
\end{equation}
where $q^m_\alpha$ is the $\rm \alpha^{th}$ member of the
set of fields defined on m-simplices in some arbitrary triangulation (an m co-chain)
and $I^{n,n-1}$ is an incidence matrix representing the
effect of the boundary operator $\delta$ acting on that
co-chain 
\begin{equation}
\delta q^n_\alpha=\sum_{j}I^{n,n-1}_{\alpha\beta}q^{n-1}_{\beta}
\end{equation}
where $I$ is only non-zero when the (n-1)-simplex lies in the boundary of the n-simplex where it takes
the value $\pm 1$ according to orientation. $\bar{I}$ is the adjoint matrix corresponding to the
co-boundary operator and gives an oriented list of (n+1)-simplices that contain the n-simplex in their boundary~\footnote{For simplicity we restrict to equilateral
simplices where each link is equal to a fixed lattice spacing.}. 
The \KD operator $K=\delta-\bar{\delta}$ is the natural
generalization of the staggered operator (thought of as a sum over neighbors in \eqref{eq:comm_full}) to a general random
triangulation suitable for describing a discretization of a curved space
\cite{Catterall:2018dns,Catterall:2018lkj}. It naturally squares to
the (discrete) Hodge Laplacian in each p-cochain sector \cite{Banks:1982iq,Catterall:2023nww}. 
It should also be noted that the fact that  the boundary of a  boundary operator $\delta^2=0$ vanishes identically is analogous to the appearance
of the minus sign for each plaquette in a hypercubic lattice. 
While the parity transformation does not exist for a general triangulation it is
possible to implement time reversal. It is given by
\begin{eqnarray}
    q_\alpha^n\stackrel{{\cal T}}{\to}\Gamma q_\alpha^n
\end{eqnarray}
where $\Gamma=\left(-1\right)^n$ is an operator that anticommutes with the \KD operator and returns $\pm 1$ depending on the order of the simplex field it acts on
(see \cite{Catterall:2018dns} for the analogous Euclidean operator).
In a regular lattice the existence of shift symmetries was related to the presence of zero modes of the staggered operator. 
On a random triangulation most of these zero modes are lifted
and the corresponding shift symmetries are lost. Nevertheless a finite number of zero modes may remain as
dictated by the index theorem
\begin{equation}
    n_+-n_-=\chi
\end{equation}
where $\chi$ is the Euler characteristic and $n_\pm$ denote the number of exact zero modes of the 
discrete K\"{a}hler-Dirac operator with eigenvalue $\pm 1$ under $\Gamma$.  This implies that there are (at least) $\chi$ conserved charges 
whose eigenvalues label states in the Hilbert space.  If we stick to the exact sequences determined by $K$, the set of zero modes is actually the cohomology of the triangulation and zero modes survive even in the odd dimensional case. These can be called $C_{\omega}$, where $\omega$ is the harmonic representative of the cohomology class. 

The supersymmetry described earlier can be generalized to these triangulations. One simply writes a supercharge 
\begin{equation}
    Q=\frac{1}{\sqrt{2}}\sum_p\sum_\alpha \theta^p_\alpha q^p_\alpha
\end{equation}
where $\alpha=1\ldots N_p$ labels the $p$-simplices in the triangulation. If we again posit that
the fermions $\theta$ satisfy anti-commutators $\{\theta^n_\alpha,\theta^m_\beta\}=2\delta^{nm}\delta_{\alpha\beta}$ then the Hamiltonian $H=Q^2$ is given
by 
\begin{equation}
    H=\frac{1}{2}\sum_p\sum_\alpha \left(q^p_\alpha\right)^2+\frac{i}{2}\theta^n_\alpha K^{n m}_{\alpha\beta}\theta^m_\beta
\end{equation}
The corresponding supersymmetry variations
\begin{equation}
    \{Q,\theta^p_\alpha\} = \sqrt 2 q^p_\alpha
\end{equation}
show that any sum over $q^p_\alpha$ that leads to a zero mode for the bosons leads to a similar zero mode for the fermions.  

When we include gauge fields, we need to think of $q, \theta$ as complex variables.  We can produce a coupling of both the bosons and the fermions to a non-trivial background gauge field by twisting the commutators of the bosons by unitaries $U$. 
Just as before, when the bosons and fermions are complex, the supersymmetry is doubled: there is a nilpotent holomorphic and an antiholomorphic $Q,\bar Q $, with $H=(Q+\bar Q)^2= (i Q-i \bar Q)^2$.

Unlike other models of Hamiltonian supersymmetry with two  supersymmetries that can be put on general graphs \cite{Fendley:2002sg},
here we have systems where the number of ground states behaves well even on large lattices.
There are a finite number of fermion zero modes that controls their degeneracy. The corresponding bosonic zero modes contribute non-trivially to the energy, as the quadratic form defining the Hamiltonian does not vanish when the bosonic zero modes are excited.

\section{Conclusion}
In this paper we have constructed Hamiltonian
lattice theories that describe supersymmetric
theories of staggered or, more generally,
\KD fields in both flat and curved space. 

In flat space the fermionic
sector of these theories correspond to staggered fermions - a well known solution
to the fermion doubling problem describing
$2^{\frac{d-1}{2}}$ flavors of degenerate Dirac fermion in the
continuum limit \cite{Susskind:1976jm, Catterall:2025vrx}. The bosonic sector is novel and arises from a non-trivial modification of the canonical
commutation relations and describes a set of p-form fields in the
continuum. These lattice theories admit
parity, time reversal and translation-by-one or shift symmetries as well
as supersymmetry. Some of these shift symmetries are expected to be non-invertible symmetries and possess non-trivial 't Hooft anomalies with parity and time reversal. These theories can also be gauged under a background $U(N)$ gauge symmetry.

This construction can also be generalized to a random triangulation which can
be thought of as representing a curved space. 
Not all symmetries
survive -- parity is lost but time reversal invariance survives as well as supersymmetry.
Ultimately, if the gauge field can be made dynamical 
one can think of these constructions as lattice analogs of the cohomological
topological quantum field theories introduced by Witten \cite{Witten:1988ze}.

Let us notice one last item. Consider a fermionic system of Majorana fermions $\theta^a$ with Hamiltonian given by
\begin{equation}
H_{fermionic} = \sum   \frac{i}{2} K_{\alpha\beta}\theta_\alpha \theta_\beta
\end{equation}
where $K_{\alpha\beta}$ is a local (antisymmetric) matrix. It is straightforward to introduce a system of staggered bosons that supersymmetrizes the system. The idea is that to each $\theta_\alpha$ we assign a $q_{\alpha}$. The commutation relations of these are
\begin{equation}
    [q_\alpha,q_\beta]= i K_{\alpha\beta}
\end{equation}
and since $K$ has to be antisymmetric, this is consistent with the bosonic statistics of $q$. 

The supercharge is given by
\begin{equation}
    Q= \frac 1{\sqrt 2} \sum_{\alpha} q_{\alpha} \theta_{\alpha}
\end{equation}
and it is easy to see that $Q^2$ produces the $H_{fermionic}$ above. Zero modes of fermions correspond to zero modes of bosons and vice-versa. In this way
we can imagine supersymmetrizing other classes of lattice fermion eg.  
domain wall fermions.

\acknowledgements

D.B. would like to thank R. Brower, P.T. Lloyd, K. Schoutens and S. Vithouladitis for discussions. D.B. would like to thank the Institute of Physics at the University of Amsterdam for their hospitality while this work was being carried out. 
The work of D.B. was supported in part by
the Department of Energy under grant DE-SC 0011702 and S.C by DE-SC-0009998.

\bibliography{shift.bib}

\begin{thebibliography}{26}%
\makeatletter
\providecommand \@ifxundefined [1]{%
 \@ifx{#1\undefined}
}%
\providecommand \@ifnum [1]{%
 \ifnum #1\expandafter \@firstoftwo
 \else \expandafter \@secondoftwo
 \fi
}%
\providecommand \@ifx [1]{%
 \ifx #1\expandafter \@firstoftwo
 \else \expandafter \@secondoftwo
 \fi
}%
\providecommand \natexlab [1]{#1}%
\providecommand \enquote  [1]{``#1''}%
\providecommand \bibnamefont  [1]{#1}%
\providecommand \bibfnamefont [1]{#1}%
\providecommand \citenamefont [1]{#1}%
\providecommand \href@noop [0]{\@secondoftwo}%
\providecommand \href [0]{\begingroup \@sanitize@url \@href}%
\providecommand \@href[1]{\@@startlink{#1}\@@href}%
\providecommand \@@href[1]{\endgroup#1\@@endlink}%
\providecommand \@sanitize@url [0]{\catcode `\\12\catcode `\$12\catcode
  `\&12\catcode `\#12\catcode `\^12\catcode `\_12\catcode `\%12\relax}%
\providecommand \@@startlink[1]{}%
\providecommand \@@endlink[0]{}%
\providecommand \url  [0]{\begingroup\@sanitize@url \@url }%
\providecommand \@url [1]{\endgroup\@href {#1}{\urlprefix }}%
\providecommand \urlprefix  [0]{URL }%
\providecommand \Eprint [0]{\href }%
\providecommand \doibase [0]{https://doi.org/}%
\providecommand \selectlanguage [0]{\@gobble}%
\providecommand \bibinfo  [0]{\@secondoftwo}%
\providecommand \bibfield  [0]{\@secondoftwo}%
\providecommand \translation [1]{[#1]}%
\providecommand \BibitemOpen [0]{}%
\providecommand \bibitemStop [0]{}%
\providecommand \bibitemNoStop [0]{.\EOS\space}%
\providecommand \EOS [0]{\spacefactor3000\relax}%
\providecommand \BibitemShut  [1]{\csname bibitem#1\endcsname}%
\let\auto@bib@innerbib\@empty
\bibitem [{\citenamefont {Kogut}\ and\ \citenamefont
  {Susskind}(1975)}]{Kogut:1974ag}%
  \BibitemOpen
  \bibfield  {author} {\bibinfo {author} {\bibfnamefont {J.~B.}\ \bibnamefont
  {Kogut}}\ and\ \bibinfo {author} {\bibfnamefont {L.}~\bibnamefont
  {Susskind}},\ }\bibfield  {title} {\bibinfo {title} {{Hamiltonian Formulation
  of Wilson's Lattice Gauge Theories}},\ }\href
  {https://doi.org/10.1103/PhysRevD.11.395} {\bibfield  {journal} {\bibinfo
  {journal} {Phys. Rev. D}\ }\textbf {\bibinfo {volume} {11}},\ \bibinfo
  {pages} {395} (\bibinfo {year} {1975})}\BibitemShut {NoStop}%
\bibitem [{\citenamefont {Catterall}\ \emph
  {et~al.}(2018{\natexlab{a}})\citenamefont {Catterall}, \citenamefont
  {Laiho},\ and\ \citenamefont {Unmuth-Yockey}}]{Catterall:2018lkj}%
  \BibitemOpen
  \bibfield  {author} {\bibinfo {author} {\bibfnamefont {S.}~\bibnamefont
  {Catterall}}, \bibinfo {author} {\bibfnamefont {J.}~\bibnamefont {Laiho}},\
  and\ \bibinfo {author} {\bibfnamefont {J.}~\bibnamefont {Unmuth-Yockey}},\
  }\bibfield  {title} {\bibinfo {title} {{Topological fermion condensates from
  anomalies}},\ }\href {https://doi.org/10.1007/JHEP10(2018)013} {\bibfield
  {journal} {\bibinfo  {journal} {JHEP}\ }\textbf {\bibinfo {volume} {10}},\
  \bibinfo {pages} {013}},\ \Eprint {https://arxiv.org/abs/1806.07845}
  {arXiv:1806.07845 [hep-lat]} \BibitemShut {NoStop}%
\bibitem [{\citenamefont {Catterall}\ \emph
  {et~al.}(2018{\natexlab{b}})\citenamefont {Catterall}, \citenamefont
  {Laiho},\ and\ \citenamefont {Unmuth-Yockey}}]{Catterall:2018dns}%
  \BibitemOpen
  \bibfield  {author} {\bibinfo {author} {\bibfnamefont {S.}~\bibnamefont
  {Catterall}}, \bibinfo {author} {\bibfnamefont {J.}~\bibnamefont {Laiho}},\
  and\ \bibinfo {author} {\bibfnamefont {J.}~\bibnamefont {Unmuth-Yockey}},\
  }\bibfield  {title} {\bibinfo {title} {{K\"ahler-Dirac fermions on Euclidean
  dynamical triangulations}},\ }\href
  {https://doi.org/10.1103/PhysRevD.98.114503} {\bibfield  {journal} {\bibinfo
  {journal} {Phys. Rev. D}\ }\textbf {\bibinfo {volume} {98}},\ \bibinfo
  {pages} {114503} (\bibinfo {year} {2018}{\natexlab{b}})},\ \Eprint
  {https://arxiv.org/abs/1810.10626} {arXiv:1810.10626 [hep-lat]} \BibitemShut
  {NoStop}%
\bibitem [{\citenamefont {Butt}\ \emph {et~al.}(2021)\citenamefont {Butt},
  \citenamefont {Catterall}, \citenamefont {Pradhan},\ and\ \citenamefont
  {Toga}}]{Butt:2021brl}%
  \BibitemOpen
  \bibfield  {author} {\bibinfo {author} {\bibfnamefont {N.}~\bibnamefont
  {Butt}}, \bibinfo {author} {\bibfnamefont {S.}~\bibnamefont {Catterall}},
  \bibinfo {author} {\bibfnamefont {A.}~\bibnamefont {Pradhan}},\ and\ \bibinfo
  {author} {\bibfnamefont {G.~C.}\ \bibnamefont {Toga}},\ }\bibfield  {title}
  {\bibinfo {title} {{Anomalies and symmetric mass generation for
  K\"ahler-Dirac fermions}},\ }\href
  {https://doi.org/10.1103/PhysRevD.104.094504} {\bibfield  {journal} {\bibinfo
   {journal} {Phys. Rev. D}\ }\textbf {\bibinfo {volume} {104}},\ \bibinfo
  {pages} {094504} (\bibinfo {year} {2021})},\ \Eprint
  {https://arxiv.org/abs/2101.01026} {arXiv:2101.01026 [hep-th]} \BibitemShut
  {NoStop}%
\bibitem [{\citenamefont {Catterall}(2023)}]{Catterall:2022jky}%
  \BibitemOpen
  \bibfield  {author} {\bibinfo {author} {\bibfnamefont {S.}~\bibnamefont
  {Catterall}},\ }\bibfield  {title} {\bibinfo {title} {{'t Hooft anomalies for
  staggered fermions}},\ }\href {https://doi.org/10.1103/PhysRevD.107.014501}
  {\bibfield  {journal} {\bibinfo  {journal} {Phys. Rev. D}\ }\textbf {\bibinfo
  {volume} {107}},\ \bibinfo {pages} {014501} (\bibinfo {year} {2023})},\
  \Eprint {https://arxiv.org/abs/2209.03828} {arXiv:2209.03828 [hep-lat]}
  \BibitemShut {NoStop}%
\bibitem [{\citenamefont {Catterall}(2024)}]{Catterall:2023nww}%
  \BibitemOpen
  \bibfield  {author} {\bibinfo {author} {\bibfnamefont {S.}~\bibnamefont
  {Catterall}},\ }\bibfield  {title} {\bibinfo {title} {{Lattice Regularization
  of Reduced K\"ahler-Dirac Fermions and Connections to Chiral Fermions}},\
  }\href {https://doi.org/10.21468/SciPostPhys.16.4.108} {\bibfield  {journal}
  {\bibinfo  {journal} {SciPost Phys.}\ }\textbf {\bibinfo {volume} {16}},\
  \bibinfo {pages} {108} (\bibinfo {year} {2024})},\ \Eprint
  {https://arxiv.org/abs/2311.02487} {arXiv:2311.02487 [hep-lat]} \BibitemShut
  {NoStop}%
\bibitem [{\citenamefont {Seiberg}\ and\ \citenamefont
  {Shao}(2024)}]{Seiberg:2023cdc}%
  \BibitemOpen
  \bibfield  {author} {\bibinfo {author} {\bibfnamefont {N.}~\bibnamefont
  {Seiberg}}\ and\ \bibinfo {author} {\bibfnamefont {S.-H.}\ \bibnamefont
  {Shao}},\ }\bibfield  {title} {\bibinfo {title} {{Majorana chain and Ising
  model - (non-invertible) translations, anomalies, and emanant symmetries}},\
  }\href {https://doi.org/10.21468/SciPostPhys.16.3.064} {\bibfield  {journal}
  {\bibinfo  {journal} {SciPost Phys.}\ }\textbf {\bibinfo {volume} {16}},\
  \bibinfo {pages} {064} (\bibinfo {year} {2024})},\ \Eprint
  {https://arxiv.org/abs/2307.02534} {arXiv:2307.02534 [cond-mat.str-el]}
  \BibitemShut {NoStop}%
\bibitem [{\citenamefont {Li}\ \emph {et~al.}(2024)\citenamefont {Li},
  \citenamefont {Wang},\ and\ \citenamefont {You}}]{Li:2024dpq}%
  \BibitemOpen
  \bibfield  {author} {\bibinfo {author} {\bibfnamefont {Y.-Y.}\ \bibnamefont
  {Li}}, \bibinfo {author} {\bibfnamefont {J.}~\bibnamefont {Wang}},\ and\
  \bibinfo {author} {\bibfnamefont {Y.-Z.}\ \bibnamefont {You}},\ }\bibfield
  {title} {\bibinfo {title} {{Quantum Many-Body Lattice C-R-T Symmetry:
  Fractionalization, Anomaly, and Symmetric Mass Generation}},\ }\href@noop {}
  {\  (\bibinfo {year} {2024})},\ \Eprint {https://arxiv.org/abs/2412.19691}
  {arXiv:2412.19691 [cond-mat.str-el]} \BibitemShut {NoStop}%
\bibitem [{\citenamefont {Gioia}\ and\ \citenamefont
  {Thorngren}(2025)}]{Gioia:2025bhl}%
  \BibitemOpen
  \bibfield  {author} {\bibinfo {author} {\bibfnamefont {L.}~\bibnamefont
  {Gioia}}\ and\ \bibinfo {author} {\bibfnamefont {R.}~\bibnamefont
  {Thorngren}},\ }\bibfield  {title} {\bibinfo {title} {{Exact Chiral
  Symmetries of 3+1D Hamiltonian Lattice Fermions}},\ }\href@noop {} {\
  (\bibinfo {year} {2025})},\ \Eprint {https://arxiv.org/abs/2503.07708}
  {arXiv:2503.07708 [cond-mat.str-el]} \BibitemShut {NoStop}%
\bibitem [{\citenamefont {Chatterjee}\ \emph {et~al.}(2025)\citenamefont
  {Chatterjee}, \citenamefont {Pace},\ and\ \citenamefont
  {Shao}}]{Chatterjee:2024gje}%
  \BibitemOpen
  \bibfield  {author} {\bibinfo {author} {\bibfnamefont {A.}~\bibnamefont
  {Chatterjee}}, \bibinfo {author} {\bibfnamefont {S.~D.}\ \bibnamefont
  {Pace}},\ and\ \bibinfo {author} {\bibfnamefont {S.-H.}\ \bibnamefont
  {Shao}},\ }\bibfield  {title} {\bibinfo {title} {{Quantized Axial Charge of
  Staggered Fermions and the Chiral Anomaly}},\ }\href
  {https://doi.org/10.1103/PhysRevLett.134.021601} {\bibfield  {journal}
  {\bibinfo  {journal} {Phys. Rev. Lett.}\ }\textbf {\bibinfo {volume} {134}},\
  \bibinfo {pages} {021601} (\bibinfo {year} {2025})},\ \Eprint
  {https://arxiv.org/abs/2409.12220} {arXiv:2409.12220 [hep-th]} \BibitemShut
  {NoStop}%
\bibitem [{\citenamefont {Pace}\ \emph {et~al.}(2025)\citenamefont {Pace},
  \citenamefont {Kim}, \citenamefont {Chatterjee},\ and\ \citenamefont
  {Shao}}]{Pace:2025rfu}%
  \BibitemOpen
  \bibfield  {author} {\bibinfo {author} {\bibfnamefont {S.~D.}\ \bibnamefont
  {Pace}}, \bibinfo {author} {\bibfnamefont {M.~L.}\ \bibnamefont {Kim}},
  \bibinfo {author} {\bibfnamefont {A.}~\bibnamefont {Chatterjee}},\ and\
  \bibinfo {author} {\bibfnamefont {S.-H.}\ \bibnamefont {Shao}},\ }\bibfield
  {title} {\bibinfo {title} {{Parity anomaly from LSM: exact valley symmetries
  on the lattice}},\ }\href@noop {} {\  (\bibinfo {year} {2025})},\ \Eprint
  {https://arxiv.org/abs/2505.04684} {arXiv:2505.04684 [cond-mat.str-el]}
  \BibitemShut {NoStop}%
\bibitem [{\citenamefont {Onogi}\ and\ \citenamefont
  {Yamaoka}(2025)}]{Onogi:2025xir}%
  \BibitemOpen
  \bibfield  {author} {\bibinfo {author} {\bibfnamefont {T.}~\bibnamefont
  {Onogi}}\ and\ \bibinfo {author} {\bibfnamefont {T.}~\bibnamefont
  {Yamaoka}},\ }\bibfield  {title} {\bibinfo {title} {{Non-singlet conserved
  charges and anomalies in 3+1 D staggered fermions}},\ }\href@noop {} {\
  (\bibinfo {year} {2025})},\ \Eprint {https://arxiv.org/abs/2509.04906}
  {arXiv:2509.04906 [hep-lat]} \BibitemShut {NoStop}%
\bibitem [{\citenamefont {Catterall}\ \emph {et~al.}(2009)\citenamefont
  {Catterall}, \citenamefont {Kaplan},\ and\ \citenamefont
  {Unsal}}]{Catterall:2009it}%
  \BibitemOpen
  \bibfield  {author} {\bibinfo {author} {\bibfnamefont {S.}~\bibnamefont
  {Catterall}}, \bibinfo {author} {\bibfnamefont {D.~B.}\ \bibnamefont
  {Kaplan}},\ and\ \bibinfo {author} {\bibfnamefont {M.}~\bibnamefont
  {Unsal}},\ }\bibfield  {title} {\bibinfo {title} {{Exact lattice
  supersymmetry}},\ }\href {https://doi.org/10.1016/j.physrep.2009.09.001}
  {\bibfield  {journal} {\bibinfo  {journal} {Phys. Rept.}\ }\textbf {\bibinfo
  {volume} {484}},\ \bibinfo {pages} {71} (\bibinfo {year} {2009})},\ \Eprint
  {https://arxiv.org/abs/0903.4881} {arXiv:0903.4881 [hep-lat]} \BibitemShut
  {NoStop}%
\bibitem [{\citenamefont {Catterall}\ \emph {et~al.}(2023)\citenamefont
  {Catterall}, \citenamefont {Giedt},\ and\ \citenamefont
  {Toga}}]{Catterall:2023tmr}%
  \BibitemOpen
  \bibfield  {author} {\bibinfo {author} {\bibfnamefont {S.}~\bibnamefont
  {Catterall}}, \bibinfo {author} {\bibfnamefont {J.}~\bibnamefont {Giedt}},\
  and\ \bibinfo {author} {\bibfnamefont {G.~C.}\ \bibnamefont {Toga}},\
  }\bibfield  {title} {\bibinfo {title} {{Holography from lattice $ \mathcal{N}
  $ = 4 super Yang-Mills}},\ }\href {https://doi.org/10.1007/JHEP08(2023)084}
  {\bibfield  {journal} {\bibinfo  {journal} {JHEP}\ }\textbf {\bibinfo
  {volume} {08}},\ \bibinfo {pages} {084}},\ \Eprint
  {https://arxiv.org/abs/2303.16025} {arXiv:2303.16025 [hep-th]} \BibitemShut
  {NoStop}%
\bibitem [{\citenamefont {Berenstein}(2023)}]{Berenstein:2023tru}%
  \BibitemOpen
  \bibfield  {author} {\bibinfo {author} {\bibfnamefont {D.}~\bibnamefont
  {Berenstein}},\ }\bibfield  {title} {\bibinfo {title} {{Staggered bosons}},\
  }\href {https://doi.org/10.1103/PhysRevD.108.074509} {\bibfield  {journal}
  {\bibinfo  {journal} {Phys. Rev. D}\ }\textbf {\bibinfo {volume} {108}},\
  \bibinfo {pages} {074509} (\bibinfo {year} {2023})},\ \Eprint
  {https://arxiv.org/abs/2303.12837} {arXiv:2303.12837 [hep-th]} \BibitemShut
  {NoStop}%
\bibitem [{\citenamefont {Berenstein}\ and\ \citenamefont
  {Lloyd}(2024)}]{Berenstein:2023ric}%
  \BibitemOpen
  \bibfield  {author} {\bibinfo {author} {\bibfnamefont {D.}~\bibnamefont
  {Berenstein}}\ and\ \bibinfo {author} {\bibfnamefont {P.~N.~T.}\ \bibnamefont
  {Lloyd}},\ }\bibfield  {title} {\bibinfo {title} {{One dimensional staggered
  bosons, clock models, and their noninvertible symmetries}},\ }\href
  {https://doi.org/10.1103/PhysRevD.110.054508} {\bibfield  {journal} {\bibinfo
   {journal} {Phys. Rev. D}\ }\textbf {\bibinfo {volume} {110}},\ \bibinfo
  {pages} {054508} (\bibinfo {year} {2024})},\ \Eprint
  {https://arxiv.org/abs/2311.00057} {arXiv:2311.00057 [hep-th]} \BibitemShut
  {NoStop}%
\bibitem [{\citenamefont {Shao}(2023)}]{Shao:2023gho}%
  \BibitemOpen
  \bibfield  {author} {\bibinfo {author} {\bibfnamefont {S.-H.}\ \bibnamefont
  {Shao}},\ }\bibfield  {title} {\bibinfo {title} {{What's Done Cannot Be
  Undone: TASI Lectures on Non-Invertible Symmetries}},\ }\href@noop {} {\
  (\bibinfo {year} {2023})},\ \Eprint {https://arxiv.org/abs/2308.00747}
  {arXiv:2308.00747 [hep-th]} \BibitemShut {NoStop}%
\bibitem [{\citenamefont {Berenstein}\ \emph {et~al.}(2024)\citenamefont
  {Berenstein}, \citenamefont {Catterall},\ and\ \citenamefont
  {Lloyd}}]{Berenstein:2024tdc}%
  \BibitemOpen
  \bibfield  {author} {\bibinfo {author} {\bibfnamefont {D.}~\bibnamefont
  {Berenstein}}, \bibinfo {author} {\bibfnamefont {S.}~\bibnamefont
  {Catterall}},\ and\ \bibinfo {author} {\bibfnamefont {P.~N.~T.}\ \bibnamefont
  {Lloyd}},\ }\bibfield  {title} {\bibinfo {title} {{Staggered bosons and
  Kahler-Dirac bosons}},\ }\href {https://doi.org/10.22323/1.463.0280}
  {\bibfield  {journal} {\bibinfo  {journal} {PoS}\ }\textbf {\bibinfo {volume}
  {CORFU2023}},\ \bibinfo {pages} {280} (\bibinfo {year} {2024})},\ \Eprint
  {https://arxiv.org/abs/2405.03758} {arXiv:2405.03758 [hep-th]} \BibitemShut
  {NoStop}%
\bibitem [{\citenamefont {Susskind}(1977)}]{Susskind:1976jm}%
  \BibitemOpen
  \bibfield  {author} {\bibinfo {author} {\bibfnamefont {L.}~\bibnamefont
  {Susskind}},\ }\bibfield  {title} {\bibinfo {title} {{Lattice Fermions}},\
  }\href {https://doi.org/10.1103/PhysRevD.16.3031} {\bibfield  {journal}
  {\bibinfo  {journal} {Phys. Rev. D}\ }\textbf {\bibinfo {volume} {16}},\
  \bibinfo {pages} {3031} (\bibinfo {year} {1977})}\BibitemShut {NoStop}%
\bibitem [{\citenamefont {Golterman}\ and\ \citenamefont
  {Smit}(1984)}]{Golterman:1984cy}%
  \BibitemOpen
  \bibfield  {author} {\bibinfo {author} {\bibfnamefont {M.~F.~L.}\
  \bibnamefont {Golterman}}\ and\ \bibinfo {author} {\bibfnamefont
  {J.}~\bibnamefont {Smit}},\ }\bibfield  {title} {\bibinfo {title}
  {{Selfenergy and Flavor Interpretation of Staggered Fermions}},\ }\href
  {https://doi.org/10.1016/0550-3213(84)90424-3} {\bibfield  {journal}
  {\bibinfo  {journal} {Nucl. Phys.}\ }\textbf {\bibinfo {volume} {B245}},\
  \bibinfo {pages} {61} (\bibinfo {year} {1984})}\BibitemShut {NoStop}%
\bibitem [{\citenamefont {Catterall}\ \emph {et~al.}(2025)\citenamefont
  {Catterall}, \citenamefont {Pradhan},\ and\ \citenamefont
  {Samlodia}}]{Catterall:2025vrx}%
  \BibitemOpen
  \bibfield  {author} {\bibinfo {author} {\bibfnamefont {S.}~\bibnamefont
  {Catterall}}, \bibinfo {author} {\bibfnamefont {A.}~\bibnamefont {Pradhan}},\
  and\ \bibinfo {author} {\bibfnamefont {A.}~\bibnamefont {Samlodia}},\
  }\bibfield  {title} {\bibinfo {title} {{Symmetries and Anomalies of
  Hamiltonian Staggered Fermions}},\ }\href@noop {} {\  (\bibinfo {year}
  {2025})},\ \Eprint {https://arxiv.org/abs/2501.10862} {arXiv:2501.10862
  [hep-lat]} \BibitemShut {NoStop}%
\bibitem [{\citenamefont {Elitzur}\ \emph {et~al.}(1982)\citenamefont
  {Elitzur}, \citenamefont {Rabinovici},\ and\ \citenamefont
  {Schwimmer}}]{Elitzur:1982vh}%
  \BibitemOpen
  \bibfield  {author} {\bibinfo {author} {\bibfnamefont {S.}~\bibnamefont
  {Elitzur}}, \bibinfo {author} {\bibfnamefont {E.}~\bibnamefont
  {Rabinovici}},\ and\ \bibinfo {author} {\bibfnamefont {A.}~\bibnamefont
  {Schwimmer}},\ }\bibfield  {title} {\bibinfo {title} {{Supersymmetric Models
  on the Lattice}},\ }\href {https://doi.org/10.1016/0370-2693(82)90269-6}
  {\bibfield  {journal} {\bibinfo  {journal} {Phys. Lett. B}\ }\textbf
  {\bibinfo {volume} {119}},\ \bibinfo {pages} {165} (\bibinfo {year}
  {1982})}\BibitemShut {NoStop}%
\bibitem [{\citenamefont {Elitzur}\ and\ \citenamefont
  {Schwimmer}(1983)}]{Elitzur:1983nj}%
  \BibitemOpen
  \bibfield  {author} {\bibinfo {author} {\bibfnamefont {S.}~\bibnamefont
  {Elitzur}}\ and\ \bibinfo {author} {\bibfnamefont {A.}~\bibnamefont
  {Schwimmer}},\ }\bibfield  {title} {\bibinfo {title} {{$N=2$ Two-dimensional
  {Wess-Zumino} Model on the Lattice}},\ }\href
  {https://doi.org/10.1016/0550-3213(83)90465-0} {\bibfield  {journal}
  {\bibinfo  {journal} {Nucl. Phys. B}\ }\textbf {\bibinfo {volume} {226}},\
  \bibinfo {pages} {109} (\bibinfo {year} {1983})}\BibitemShut {NoStop}%
\bibitem [{\citenamefont {Banks}\ \emph {et~al.}(1982)\citenamefont {Banks},
  \citenamefont {Dothan},\ and\ \citenamefont {Horn}}]{Banks:1982iq}%
  \BibitemOpen
  \bibfield  {author} {\bibinfo {author} {\bibfnamefont {T.}~\bibnamefont
  {Banks}}, \bibinfo {author} {\bibfnamefont {Y.}~\bibnamefont {Dothan}},\ and\
  \bibinfo {author} {\bibfnamefont {D.}~\bibnamefont {Horn}},\ }\bibfield
  {title} {\bibinfo {title} {{Geometric Fermions}},\ }\href
  {https://doi.org/10.1016/0370-2693(82)90571-8} {\bibfield  {journal}
  {\bibinfo  {journal} {Phys. Lett.}\ }\textbf {\bibinfo {volume} {B117}},\
  \bibinfo {pages} {413} (\bibinfo {year} {1982})}\BibitemShut {NoStop}%
\bibitem [{\citenamefont {Fendley}\ \emph {et~al.}(2003)\citenamefont
  {Fendley}, \citenamefont {Schoutens},\ and\ \citenamefont
  {de~Boer}}]{Fendley:2002sg}%
  \BibitemOpen
  \bibfield  {author} {\bibinfo {author} {\bibfnamefont {P.}~\bibnamefont
  {Fendley}}, \bibinfo {author} {\bibfnamefont {K.}~\bibnamefont {Schoutens}},\
  and\ \bibinfo {author} {\bibfnamefont {J.}~\bibnamefont {de~Boer}},\
  }\bibfield  {title} {\bibinfo {title} {{Lattice models with N=2
  supersymmetry}},\ }\href {https://doi.org/10.1103/PhysRevLett.90.120402}
  {\bibfield  {journal} {\bibinfo  {journal} {Phys. Rev. Lett.}\ }\textbf
  {\bibinfo {volume} {90}},\ \bibinfo {pages} {120402} (\bibinfo {year}
  {2003})},\ \Eprint {https://arxiv.org/abs/hep-th/0210161}
  {arXiv:hep-th/0210161} \BibitemShut {NoStop}%
\bibitem [{\citenamefont {Witten}(1988)}]{Witten:1988ze}%
  \BibitemOpen
  \bibfield  {author} {\bibinfo {author} {\bibfnamefont {E.}~\bibnamefont
  {Witten}},\ }\bibfield  {title} {\bibinfo {title} {{Topological Quantum Field
  Theory}},\ }\href {https://doi.org/10.1007/BF01223371} {\bibfield  {journal}
  {\bibinfo  {journal} {Commun. Math. Phys.}\ }\textbf {\bibinfo {volume}
  {117}},\ \bibinfo {pages} {353} (\bibinfo {year} {1988})}\BibitemShut
  {NoStop}%
\end{thebibliography}%

\end{document}